\documentclass[aps,prb,twocolumn,showpacs,floatfix,longbibliography]{revtex4-1}
\usepackage{graphicx,amsfonts,amssymb,amsmath}
\usepackage{textcomp}
\usepackage{esint}
\usepackage[colorlinks,linktocpage,bookmarks=false,citecolor=blue,linkcolor=red,urlcolor=blue]{hyperref}
\usepackage[utf8]{inputenc}
\usepackage[dvipsnames]{xcolor}
\usepackage[english]{babel}
\usepackage{afterpage}
\usepackage{bbold}

\input cyracc.def
\font\tencyr=wncysc10
\def\cyr{\tencyr\cyracc}
\def\sh{\mbox{\cyr SH}}

\allowdisplaybreaks

\def\phiset{\{\phi_{ia}\}}
\def\varphiset{\{\varphi_{ia}\}}
\def\Jset{\{J_{ia}\}}
\def\phipset{\{\phi'_{ia}\}}

\def\Jfset{\{J_{jf}\}} 
 
\def\xiov{\xi_{\rm ov}} 
\def\nmax{n_{\rm max}}

\begin{document}


\def\rhoeq{\hat\rho_{\rm eq}}

\newcommand{\marge}[1]{\marginpar{\scriptsize #1}}
\newcommand{\remarque}[1]{\marginpar{\scriptsize Remarque}{\it [#1]}}
\newcommand{\new}[1]{{\bf #1}}
\newcommand{\red}[1]{\textcolor{red}{#1}}
\newlength{\textlarg}
\newcommand{\redbar}[1]{\textcolor{red}{\st{#1}}} 
\newcommand{\bluebar}[1]{\textcolor{blue}{\st{#1}}} 

\newcommand{\beq}{\begin{equation}}
\newcommand{\eeq}{\end{equation}}
\newcommand{\bfig}{\begin{figure}}
\newcommand{\efig}{\end{figure}}
\newcommand{\bline}{\begin{multline}}
\newcommand{\eline}{\end{multline}}
\newcommand{\bremark}{\begin{quotation} \noindent \small }
\newcommand{\eremark}{\end{quotation}}
\newcommand{\llbrace}{\left\lbrace}  
\newcommand{\rrbrace}{\right\rbrace}
\newcommand{\lbraket}{\left[}
\newcommand{\rbraket}{\right]}
\newcommand{\llangle}{\left\langle}
\newcommand{\rrangle}{\right\rangle} 

\newcommand{\Tr}{{\rm Tr}} 
\newcommand{\tr}{{\rm tr}} 
\newcommand{\sgn}{\,{\rm sgn}} 
\newcommand{\mean}[1]{\langle #1 \rangle}
\newcommand{\commu}[2]{[#1,#2]} 
\newcommand{\bra}[1]{\langle#1|}
\newcommand{\ket}[1]{|#1\rangle}
\newcommand{\braket}[2]{\langle #1|#2\rangle}
\newcommand{\ketbra}[2]{|#1\rangle\langle#2|}
\newcommand{\dbraket}[3]{\langle #1|#2|#3\rangle}
\newcommand{\tens}[1]{\overleftrightarrow{#1}}  
\newcommand{\vac}{|{\rm vac}\rangle} 
\newcommand{\bravac}{\langle{\rm vac}|}
\newcommand{\const}{{\rm const}} 
\newcommand{\unif}{{\rm unif.}} 
\newcommand{\atanh}{\,{\rm atanh}}
\newcommand{\cotanh}{\,{\rm cotanh}}

\newcommand{\ie}{i.e.\xspace}
\newcommand{\iet}{i.e.}
\newcommand{\eg}{e.g.\xspace}
\newcommand{\cc}{{\rm c.c.}} 
\newcommand{\hc}{{\rm h.c.}} 
\newcommand{\etal}{{\it et al. }}
\newcommand\eme{$^{\mbox{\footnotesize ème}}$\xspace}

\newcommand{\jhatbf}{\hat {\textbf \jold}} 
\newcommand{\Jhatbf}{\hat {\textbf \J}} 
\newcommand{\jhat}{\hat {\jmath}} 
\newcommand{\Jhat}{\hat {J}} 
\newcommand{\jbf}{\textbf j}
\newcommand{\Jbf}{\textbf J}

\def\chibf{\boldsymbol{\chi}}
\def\down{\downarrow}
\def\eps{\epsilon}
\def\gam{\gamma} 
\def\alphabf{\boldsymbol{\alpha}}
\def\phibf{\boldsymbol{\phi}}
\def\varphibf{\boldsymbol{\varphi}}
\def\varphibfs{\boldsymbol{\varphi}_<}
\def\varphibfl{\boldsymbol{\varphi}_>}
\def\varphis{\varphi_{<}}
\def\varphil{\varphi_{>}}
\def\psibf{\boldsymbol{\psi}}
\def\thetabf{\boldsymbol{\theta}}
\def\Ome{\Omega}
\def\omeD{{\omega_D}} 
\def\bfOme{\boldsymbol{\Omega}} 
\def\Omebf{\boldsymbol{\Omega}} 
\def\lamb{\lambda}
\def\Lamb{\Lambda}
\def\sig{\sigma}
\def\Sig{\Sigma}
\def\sigp{{\sigma'}} 
\def\bfsig{\boldsymbol{\sigma}} 
\def\sigbf{\boldsymbol{\sigma}} 
\def\bfSig{\boldsymbol{\Sigma}} 
\def\The{\Theta} 
\def\up{\uparrow}

\def\epsk{\epsilon_{\bf k}} 
\def\xik{\xi_{\bf k}} 
\def\txik{\tilde\xi_{\bf k}} 
\def\xip{\xi_{\bf p}} 
\def\xiq{\xi_{\bf q}} 
\def\xikq{\xi_{{\bf k}+{\bf q}}} 
\def\Ek{E_{\bf k}} 
\def\Ep{E_{\bf p}}
\def\Eq{E_{\bf q}}
\def\Heff{\hat H_{\rm eff}}
\def\Hem{\hat H_{\rm em}}
\def\Hint{\hat H_{\rm int}}
\def\Hloc{\hat H_{\rm loc}}
\def\HMF{\hat H_{\rm MF}}
\def\HLL{\hat H_{\rm LL}}
\def\Sem{S_{\rm em}}
\def\SMF{S_{\rm MF}} 
\def\SHF{S_{\rm HF}} 
\def\SRPA{S_{\rm RPA}} 
\def\Sint{S_{\rm int}} 
\def\Sloc{S_{\rm loc}}
\def\TN{T_{\rm N}} 
\def\TNHF{T^{\rm HF}_{\rm N}} 
\def\Zloc{Z_{\rm loc}} 
\def\ZMF{Z_{\rm MF}} 
\def\ZHF{Z_{\rm HF}} 
\def\ZRPA{Z_{\rm RPA}} 
\def\RPA{{\rm RPA}}
\def\loc{{\rm loc}} 
\def\pp{{\rm pp}}
\def\ph{{\rm ph}} 
\def\ch{{\rm ch}}
\def\sp{{\rm sp}} 
\def\qtf{q_{\rm TF}}
\def\epstf{\eps^{}_{\rm TF}} 
\def\epsrpa{\eps^{}_{\rm RPA}} 
\def\chinnzpp{\chi_{nn}^{0}{}\!\!\!''}

\def\half{\frac{1}{2}}
\def\dhalf{\dfrac{1}{2}}
\def\third{\frac{1}{3}} 
\def\quarter{\frac{1}{4}}

\def\qr{{\bf q}\cdot{\bf r}}
\def\wt{\omega t} 

\def\a{{\bf a}}
\def\b{{\bf b}}
\newcommand{\cv}{{\bf c}} 
\def\e{{\bf e}}
\def\f{{\bf f}}
\def\g{{\bf g}}
\def\h{{\bf h}}
\def\jold{\char"11}
\def\j{{\bf j}}
\def\k{{\bf k}}
\def\l{{\bf l}}
\def\m{{\bf m}}
\def\n{{\bf n}} 
\def\p{{\bf p}} 
\def\q{{\bf q}}
\def\r{{\bf r}}
\def\t{{\bf t}}
\def\u{{\bf u}}
\newcommand{\vv}{{\bf v}}
\def\x{{\bf x}}
\def\y{{\bf y}} 
\def\z{{\bf z}} 
\def\A{{\bf A}}
\def\B{{\bf B}}
\def\D{{\bf D}} 
\def\E{{\bf E}} 
\def\F{{\bf F}} 
\def\H{{\bf H}}  
\def\J{{\bf J}}
\def\K{{\bf K}} 

\def\G{{\bf G}}
\def\L{{\bf L}}
\def\M{{\bf M}}  
\def\O{{\bf O}} 
\def\P{{\bf P}} 
\def\Q{{\bf Q}} 
\def\R{{\bf R}}
\def\S{{\bf S}}
\def\U{{\bf U}} 
\def\V{{\bf V}} 
\def\X{{\bf X}} 
\def\Y{{\bf Y}} 
\def\epsbf{\boldsymbol{\epsilon}}
\def\betabf{\boldsymbol{\beta}}
\def\deltabf{\boldsymbol{\delta}}
\def\mubf{\boldsymbol{\mu}}
\def\nablabf{\boldsymbol{\nabla}}
\def\rhobf{\boldsymbol{\rho}}
\def\sigmabf{\boldsymbol{\sigma}} 
\def\Pibf{\boldsymbol{\Pi}}
\def\pibf{\boldsymbol{\pi}}

\def\para{\parallel}
\def\kpara{{k_\parallel}}
\def\kperp{{k_\perp}} 
\def\kperpp{{k_\perp'}} 
\def\qperp{{q_\perp}} 
\def\tperp{{t_\perp}} 

\def\w{\omega}
\def\wn{\omega_n}
\def\wm{\omega_m}
\def\wnu{\omega_\nu}
\def\wp{\omega_p} 
\def\dmu{{\partial_\mu}}
\def\dnu{{\partial_\nu}}
\def\dl{{\partial_l}}  
\def\dt{\partial_t} 
\def\tdt{\tilde\partial_t}
\def\dk{\partial_k}
\def\tdk{\tilde\partial_k}
\def\dx{\partial_x}
\def\dy{\partial_y} 
\def\dtau{{\partial_\tau}}  
\def\det{{\rm det}} 
\def\Pf{{\rm Pf}}
\def\diag{{\rm diag}}

\def\dsum{\displaystyle \sum}
\def\dint{\displaystyle \int} 
\def\intt{\int_{-\infty}^\infty dt} 
\def\inttp{\int_{-\infty}^\infty dt'} 
\def\intk{\int_{\bf k}} 
\def\intkd{\int \frac{d^dk}{(2\pi)^d}}
\def\intq{\int_{\bf q}} 
\def\intr{\int d^dr}  
\def\dintr{\displaystyle \int d^dr} 
\def\intrp{\int d^dr'}
\def\dinttau{\displaystyle \int_0^\beta d\tau}
\def\dinttaup{\displaystyle \int_0^\beta d\tau'}
\def\inttau{\int_0^\beta d\tau}
\def\inttaup{\int_0^\beta d\tau'}
\def\intx{\int d^{d+1}x} 
\def\inttaur{\int_0^\beta d\tau \int d^dr}
\def\intinf{\int_{-\infty}^\infty}
\def\dinttaur{\displaystyle \int_0^\beta d\tau \int d^dr}
\def\dintinf{\displaystyle \int_{-\infty}^\infty}
\def\intw{\int_{-\infty}^\infty \frac{d\w}{2\pi}}
\def\sumr{\sum_{\bf r}} 

\def\calA{{\cal A}}
\def\calAbf{\bm{{\cal A}}}
\def\calB{{\cal B}} 
\def\calC{{\cal C}} 
\def\dt{\partial_t}
\def\calD{{\cal D}}
\def\calE{{\cal E}}
\def\calF{{\cal F}} 
\def\calFbf{\bm{{\cal F}}}
\def\calG{{\cal G}}
\def\calH{{\cal H}}
\def\calI{{\cal I}}
\def\calJ{{\cal J}}
\def\calK{{\cal K}}
\def\calL{{\cal L}} 
\def\calM{{\cal M}} 
\def\calN{{\cal N}}
\def\calO{{\cal O}}
\def\calP{{\cal P}}  
\def\calR{{\cal R}} 
\def\calS{{\cal S}}
\def\calT{{\cal T}}
\def\calU{{\cal U}}
\def\calV{{\cal V}}
\def\calX{{\cal X}} 
\def\calY{{\cal Y}} 
\def\calZ{{\cal Z}} 

\def\calbfB{{\bf \cal B}}
\def\calbfF{{\bf \cal F}}

\def\tT{{\tilde T}}
\def\talpha{{\tilde\alpha}}
\def\tbeta{{\tilde\beta}}
\def\tchi{{\tilde\chi}}
\def\tdelta{{\tilde\delta}}
\def\tDelta{{\tilde\Delta}}
\def\teta{{\tilde\eta}} 
\def\tlamb{{\tilde\lambda}}
\def\tmu{{\tilde\mu}}
\def\tphibf{{\tilde\phibf}}
\def\trho{{\tilde\rho}}
\def\tvarphibf{{\tilde\varphibf}} 
\def\tw{{\tilde\omega}}
\def\twn{{\tilde\omega_n}}
\def\twnu{{\tilde\omega_\nu}}

\def\asinh{{\rm asinh}} 
\def\Tbkt{T_{\rm BKT}}

\graphicspath{{./figures/}}

\title{Chaos in the Bose-glass phase of a one-dimensional disordered Bose fluid}

\author{Romain Daviet}
\author{Nicolas Dupuis}
\affiliation{Sorbonne Universit\'e, CNRS, Laboratoire de Physique Th\'eorique de la Mati\`ere Condens\'ee, LPTMC, F-75005 Paris, France}

\date{April 27, 2021} 

\begin{abstract}
We show that the Bose-glass phase of a one-dimensional disordered Bose fluid exhibits a chaotic behavior, i.e., an extreme sensitivity to external parameters. Using bosonization, the replica formalism and the nonperturbative functional renormalization group, we find that the ground state is unstable to any modification of the disorder configuration (``disorder'' chaos) or variation of the Luttinger parameter (``quantum'' chaos, analog to the ``temperature'' chaos in classical disordered systems). This result is obtained by considering two copies of the system, with slightly different disorder configurations or Luttinger parameters, and showing that inter-copy statistical correlations are suppressed at length scales larger than an overlap length $\xiov\sim |\eps|^{-1/\alpha}$ ($|\eps|\ll 1$ is a measure of the difference between the disorder distributions or Luttinger parameters of the two copies). The chaos exponent $\alpha$ can be obtained by computing $\xiov$ or by studying the instability of the Bose-glass fixed point for the two-copy system when $\eps\neq 0$. The renormalized, functional, inter-copy disorder correlator departs from its fixed-point value -- characterized by cuspy singularities -- {\it via} a chaos boundary layer, in the same way as it approaches the Bose-glass fixed point when $\eps=0$ through a quantum boundary layer. Performing a linear analysis of perturbations about the Bose-glass fixed point, we find $\alpha=1$. 
\end{abstract}

\maketitle

\tableofcontents

\section{Introduction}

In a Bose fluid with short-range interactions, disorder can induce a quantum phase transition between a superfluid phase and a localized phase dubbed Bose glass (BG).\cite{Giamarchi87,Giamarchi88,Fisher89} The latter is characterized by a nonzero compressibility, a vanishing dc conductivity and the absence of gap in the optical conductivity. As its name indicates, the BG phase is expected to be analogous to the Fermi-glass phase of interacting fermions in a strong disorder potential and exhibit some of the characteristic properties of glassy systems.\cite{Fisher89} 

In one dimension, the analogy of the BG phase with other disordered systems exhibiting glassy properties is strongly supported by the nonperturbative functional renormalization group (FRG).\cite{Dupuis19,Dupuis20,Dupuis20a,Daviet20} In this approach, one finds that the BG phase is described by an attractive fixed point analog to the zero-temperature fixed point controlling the low-temperature phase of many classical disordered systems. The role of temperature is played by the Luttinger parameter $K\sim k^\theta$ which, as the momentum scale $k$ approaches zero, vanishes with an exponent $\theta=z-1$ related to the dynamical critical exponent $z$. Moreover, the renormalized disorder correlator assumes a cuspy functional form associated with the existence of metastable states.\cite{Balents96} At nonzero momentum scale, quantum tunneling between the ground state and these metastable states leads to a rounding of the cusp singularity into a quantum boundary layer (QBL). The latter controls the low-energy dynamics and is responsible for the $\w^2$ behavior of the (dissipative) conductivity. Thus the FRG approach reveals some of the glassy properties (pinning, ``shocks'' or static avalanches) of the BG phase and, to some extent, can be understood within the “droplet” picture\cite{Fisher88b} put forward for the description of glassy (classical) systems.\cite{Dupuis20} 

One of the peculiar features of glassy systems is the extreme sensitivity of the ground state with respect to small changes in external parameters like the disorder configuration or the temperature. In some cases, an infinitesimal perturbation is sufficient to lead to a complete reorganization of the ground state at large length scales. This situation is referred to as chaos,\cite{McKay82,Bray87,Fisher88a,Fisher91,Shapir91,Kondor93,Kisker98} e.g. disorder chaos or temperature chaos according to the external parameter being considered. Chaos is usually characterized by an overlap length $\xiov$ beyond which the ground state completely changes as a result of the variation in the external parameter. The overlap length diverges as $\xiov\sim |\eps|^{-1/\alpha}$ where $\alpha$ is called the chaos exponent ($|\eps|\ll 1$ is a measure of the change in the external parameter). Although chaos was originally predicted for spin glasses, it is also characteristic of elastic manifolds pinned by disorder where the long-distance physics is controlled by a zero-temperature fixed point.\cite{Shapir91,Ledoussal06,Duemmer07} To our knowledge, the only quantum disordered system where (disorder) chaos was studied is the two-dimensional Anderson insulator.\cite{Lemarie19}

In this paper, we study chaos in the BG phase of a one-dimensional Bose fluid. In Sec.~\ref{sec_model} we briefly recall the FRG formalism used in Refs.~\onlinecite{Dupuis19,Dupuis20} to study the BG phase and generalize it to include two copies of the system subjected to slightly different disorder configurations.\cite{not6} In particular we introduce the main quantities of interest: the running Luttinger parameter $K_k$ and the intra- and inter-copy renormalized disorder correlators, $\delta_{11,k}(u)=\delta_{22,k}(u)$ and $\delta_{12,k}(u)$ respectively. Here $k$ is a running momentum scale and $u\equiv\phi_a-\phi_b$ stands for the difference between the fields in two different replicas. The flow equations for $K_k$ and $\delta_{ij,k}(u)$ are similar to those obtained  for pinned disordered periodic manifolds by Duemmer and Le Doussal (DLD) in Refs.~\onlinecite{Ledoussal06,Duemmer07}, with $K_k$ playing the role of the temperature.

In section~\ref{subsec_approach} we first consider the approach to the BG fixed point when the two copies are identical (i.e., experience the same disorder potential: $\eps=0$). In that case, the $\pi$-periodic functions $\delta_{ii,k}(u)$ and $\delta_{12,k}(u)$ approach a fixed-point function $\delta^*(u)$ exhibiting cusps at $u=p\pi$ ($p$ integer). At nonzero momentum scale the cusp singularity at $u=p\pi$ is rounded into a QBL. A linear analysis of the perturbations about $\delta^*(u)$ shows that the less irrelevant eigenvalue $\lamb=-\theta$ is associated with an eigenfunction which is increasingly peaked around $u=p\pi$ as the number $\nmax$ of circular harmonics of $\delta_{ij,k}(u)$ (used in the numerical solution of the linearized flow equations) increases, whereas all other eigenfunctions remain extended over the whole interval $[0,\pi]$.  

In Sec.~\ref{subsec_escape} we show that the BG fixed point is unstable for any nonzero $\eps$ since in that case $\delta_{ii,k}(u)\to \delta^*(u)$ but $\delta_{12,k}(u)\to 0$ in the limit $k\to 0$. Thus the two copies become statistically independent in the large-distance limit, which corresponds to disorder chaos. From the numerical solution of the flow equations we find that $\delta_{12,k}(0)$ satisfies a scaling form with a characteristic length $\xiov\sim|\eps|^{-1/\alpha}$ but the chaos exponent $\alpha$ seems to converge very slowly with $\nmax$. The instability of the BG fixed point occurs {\it via} a chaos boundary layer\cite{Ledoussal06,Duemmer07} (CBL) reminiscent of the QBL observed in the approach to the BG fixed point. The linear analysis of the perturbations about the fixed-point solution $\delta^*(u)$ reveals a single positive eigenvalue $\lamb\equiv 2\alpha$ associated with a function which is increasingly peaked around $u=p\pi$ as $\nmax$ increases. The convergence of $\alpha$ with $\nmax$ is extremely slow but in the limit $\nmax\to\infty$ the solution can be found analytically (and is essentially given by a Dirac comb) and yields the chaos exponent $\alpha=1$. We are then able to show that the convergence of $\alpha$ with $\nmax$ is logarithmic. The agreements and differences between our results and those of DLD are discussed in Sec.~\ref{subsubsec_DLD}. Finally, in Sec.~\ref{sec_qchaos}, we show that chaos is also obtained when one considers a slight change in the Luttinger parameter.

\section{Model and FRG formalism} 
\label{sec_model}

We consider a one-dimensional Bose fluid described by the Hamiltonian $\hat H_0$. At low energies $\hat H_0$ can be approximated by the Tomonaga-Luttinger Hamiltonian\cite{Giamarchi_book,Haldane81,Cazalilla11}
\begin{equation}
\hat H_0 = \int dx \frac{v}{2\pi} \left\{ \frac{1}{K} (\dx \hat\varphi)^2 + K (\dx \hat\theta)^2 \right\} , 
\label{H0}
\end{equation}
where $\hat\theta$ is the phase of the boson operator $\hat\psi(x)=e^{i\hat\theta(x)}\hat\rho(x)^{1/2}$ and $\hat\varphi$ is related to the density operator {\it via} 
\beq 
\hat\rho(x) =\rho_0 - \frac{1}{\pi}\dx\hat\varphi(x) 
+ 2\rho_2 \cos(2\pi\rho_0 x-2\hat\varphi(x)) , 
\eeq 
where $\rho_0$ is the average density and $\rho_2$ a nonuniversal parameter that depends on microscopic details. $\hat\varphi$ and $\hat\theta$ satisfy the commutation relations $[\hat\theta(x),\partial_y\hat\varphi(y)]=i\pi\delta(x-y)$. $v$ denotes the sound-mode velocity and the dimensionless parameter $K$, which encodes the strength of boson-boson interactions, is the Luttinger parameter. The ground state of $\hat H_0$ is a Luttinger liquid, i.e., a superfluid state with superfluid stiffness $\rho_s=vK/\pi$ and compressibility $\kappa=d\rho_0/d\mu=K/\pi v$.\cite{Giamarchi_book}

The disorder contributes to the Hamiltonian a term\cite{Giamarchi87,Giamarchi88}  
\begin{equation}
\hat H_{\rm dis} = \int dx \left\{ - \frac{1}{\pi} \eta \dx \hat\varphi + \rho_2 [ \xi^* e^{2i\hat\varphi} + \hc ] \right\} , 
\label{Hdis} 
\end{equation}
where $\eta(x)$ (real) and $\xi(x)$ (complex) denote random potentials with Fourier components near 0 and $\pm 2\pi\rho_0$, respectively. $\eta$ can be eliminated by a shift of $\hat\varphi$
and is not considered in the following.\cite{Dupuis20}

In the functional-integral formalism, after integrating out the field $\theta$, one obtains the Euclidean (imaginary-time) action
\begin{align}
S[\varphi;\xi] ={}& \int_{X} \biggl\{ \frac{v}{2\pi K} \left[ (\dx\varphi)^2 + v^{-2} (\dtau \varphi)^2 \right] \nonumber \\ & 
+ \rho_2 [ \xi^* e^{2i\varphi} + \cc ] \biggr\} ,
\label{action0}
\end{align}
where we use the notation $X=(x,\tau)$, $\int_X=\inttau\int dx$ and $\varphi(X)$ is a bosonic field with $\tau\in[0,\beta]$. The model is regularized by a UV cutoff $\Lambda$ acting both on momenta and frequencies. We shall only consider the zero-temperature limit $\beta=1/T\to\infty$ but $\beta$ will be kept finite at intermediate stages of calculations.

\subsection{Introducing two copies and $n$ replicas}

To investigate the chaotic nature of the BG phase, we consider two copies of the system with slightly different realizations of the disorder, 
\beq 
\begin{split}
\xi_1(x) &= \xi(x) + \eps \zeta(x) , \\ 
\xi_2(x) &= \xi(x) - \eps \zeta(x) ,
\end{split}
\eeq 
where $|\eps|\ll 1$. The random potentials $\xi$ and $\zeta$ are uncorrelated and identically distributed, i.e., assuming Gaussian distributions with zero mean,  
\beq 
\begin{split}
&\overline{\xi(x)} = \overline{\zeta(x)} = 0 , \\ 
&\overline{\xi^*(x)\xi(x')} = \overline{\zeta^*(x)\zeta(x')} = D\delta(x-x') 
\end{split}
\label{xiaverage} 
\eeq 
(all other correlators, e.g. $\overline{\xi(x)\xi(x')}$, vanish). We use an overline to denote disorder averaging. Equations~(\ref{xiaverage}) imply
\beq 
\begin{split}
&\overline{\xi^*_i(x) \xi_j(x')} = D_{ij} \delta(x-x') , \\ 
&D_{ii} = D(1+\eps^2) , \quad D_{12}=D_{21} = D(1-\eps^2) .
\end{split}
\label{dis}
\eeq 
The statistical correlations between the two systems are characterized by the correlation functions 
\beq 
C_{ij}(X-X') = \overline{\mean{(\varphi_i(X)-\varphi_i(X'))(\varphi_j(X)-\varphi_j(X'))}} .
\eeq 
Since the two copies are independent before disorder averaging, 
\beq 
\begin{split}
C_{ii}(X-X') ={}& 2 [ G_{c,ii}(0) + G_{d,ii}(0) \\ 
& - G_{c,ii}(X-X') - G_{d,ii}(X-X') ] ,  \\	
C_{12}(X-X') ={}& 2 [ G_{d,12}(0) - G_{d,12}(X-X') ] , 
\end{split}
\eeq 
where 
\beq 
\begin{split}
G_{c,ij}(X-X') &= \overline{\mean{\varphi_i(X) \varphi_j(X')}} - \overline{\mean{\varphi_i(X)} \mean{\varphi_j(X')}} , \\
G_{d,ij}(X-X') &= \overline{\mean{\varphi_i(X)} \mean{\varphi_j(X')}} - \overline{\mean{\varphi_i(X)}} \; \overline{\mean{\varphi_j(X')}}
\end{split}
\eeq 
are the connected and disconnected propagators, respectively. The long-distance part of both $C_{ii}(X-X')$ and $C_{12}(X-X')$ is determined by $G_{d,ij}(X-X')$.\cite{Dupuis20}

In the replica formalism, one considers $n$ replicas of the system and the disorder-averaged partition function 
\beq 
\calZ[\{J_{ia}\}] = \overline{ \prod_{a=1}^n \prod_{i=1}^2 \calZ[J_{ia};\xi_i] } ,
\eeq 
where the $2n$ external sources $\{J_{ia}\}$ act on each replica independently and 
\beq 
\calZ[J_{ia};\xi_i] = \int \calD[\varphi_{ia}] \, e^{-S[\varphi_{ia},\xi_i] + \int_X J_{ia}(X)\varphi_{ia}(X)}
\eeq 
is the partition function of the $a$th replica of the $i$th copy before disorder averaging. Using~(\ref{dis}) to perform the disorder average, one obtains  
\beq 
\calZ[\{J_{ia}\}] = \int \calD[\{\varphi_{ia}\}]\, e^{-S[\{\varphi_{ia}\}] + \sum_{a,i} \int_X J_{ia}(X)\varphi_{ia}(X) } , 
\eeq 
with the replicated action 
\begin{multline}
S[\{\varphi_{ia}\}] = \sum_{i,a} \int_{x,\tau} \frac{v}{2\pi K} \left\{ (\dx\varphi_{ia})^2 + \frac{(\dtau\varphi_{ia})^2}{v^2} \right\} \\
- \sum_{a,b,i,j} \calD_{ij} \int_{x,\tau,\tau'} \cos[2\varphi_{ia}(x,\tau) - 2\varphi_{jb}(x,\tau')] , 
\label{action} 
\end{multline} 
where $\calD_{ij}=\rho_2^2D_{ij}$.

\subsection{Effective action and FRG} 

To implement the nonperturbative FRG approach,\cite{Berges02,Kopietz_book,Delamotte12,Dupuis_review} we add to the action~(\ref{action}) the infrared regulator term\cite{Dupuis20}
\begin{equation}
\Delta S_k[\varphiset] = \half \sum_{i,a,q,\w} \varphi_{ia}(-q,-i\w) R_{k}(q,i\w) \varphi_{ia}(q,i\w) ,
\label{DeltaSk} 
\end{equation}
where $k$ is a (running) momentum scale varying from the UV scale $\Lambda$ down to zero and $\w\equiv\wn=2\pi n/\beta$ ($n$ integer) is a Matsubara frequency. The cutoff function $R_k(q,i\w)$ is chosen so that fluctuation modes satisfying $|q|,|\w|/v_k\ll k$ are suppressed while those with $|q|\gg k$ or $|\w|/v_k\gg k$ are left unaffected (the $k$-dependent sound-mode velocity $v_k$ is defined below). In practice we choose
\beq 
R_k(q,i\w) = Z_x \left(q^2 + \frac{\w^2}{v_k^2} \right) r\left( \frac{q^2+\w^2/v_k^2}{k^2} \right) , 
\eeq
where $r(y)=\alpha/(e^y-1)$ with $\alpha$ a constant of order unity. $Z_x$ is defined below. 

The partition function
\begin{align}
\calZ_k[\Jset] ={}& \int \calD[\varphiset] \exp \Bigl\{ -S[\varphiset] \nonumber \\ & 
 - \Delta S_k[\varphiset]+ \sum_{i,a} \int_{X} J_{ia}\varphi_{ia} \Bigr\} 
\end{align}
thus becomes $k$ dependent. The expectation value of the field reads 
\beq
\phi_{ia}(X) = \frac{\delta\ln \calZ_k[\Jfset]}{\delta J_{ia}(X)} =\mean{\varphi_{ia}(X)} 
\eeq 
(to avoid confusion in the indices we denote by $\Jfset$ the $2n$ external sources). 

The scale-dependent effective action 
\beq
\Gamma_k[\phiset] = - \ln \calZ_k[\Jset] + \sum_{i,a} \int_{X} J_{ia} \phi_{ia} - \Delta S_k[\phiset]
\eeq
is defined as a modified Legendre transform which includes the subtraction of $\Delta S_k[\phiset]$. Assuming that for $k=\Lamb$ the fluctuations are completely frozen by the term $\Delta S_{\Lamb}$, $\Gamma_{\Lamb}[\phiset]=S[\phiset]$. On the other hand the effective action of the original model~(\ref{action}) is given by $\Gamma_{k=0}$ since $R_{k=0}$ vanishes. The nonperturbative FRG approach aims at determining $\Gamma_{k=0}$ from $\Gamma_{\Lamb}$ using Wetterich's equation\cite{Wetterich93,Ellwanger94,Morris94} 
\beq
\dt \Gamma_k[\phiset] = \half \Tr \left\{ \dt R_k \bigl(\Gamma_k^{(2)}[\phiset] + R_k \bigr)^{-1} \right\} ,
\label{eqwet}
\eeq
where $\Gamma_k^{(2)}$ is the second functional derivative of $\Gamma_k$ and $t=\ln(k/\Lamb)$ a (negative) RG ``time''. The trace in~(\ref{eqwet}) involves a sum over momenta and frequencies as well as copy and replica indices. 

To solve (approximately) the flow equation~(\ref{eqwet}) we consider the following ansatz for the effective action\cite{Dupuis19,Dupuis20} 
\beq 
\Gamma_k[\phiset] = \sum_a \Gamma_{1,k}[\phi_{a}] - \half \sum_{a,b}\Gamma_{2,k}[\phi_{a},\phi_{b}] ,
\label{ansatz1}
\eeq 
where $\phi_a=\{\phi_{1a},\phi_{2a}\}$ and 
\begin{align} 
&\Gamma_{1,k}[\phi_{a}] = \sum_i \int_X \frac{Z_x}{2} \llbrace (\dx \phi_{ia})^2 + \frac{(\dtau\phi_{ia})^2}{v_k^2}  \rrbrace , \nonumber \\ 
&\Gamma_{2,k}[\phi_a,\phi_b] = \sum_{i,j} \int_{x,\tau,\tau'} V_{ij,k} \bigl(\phi_{ia}(x,\tau) - \phi_{jb}(x,\tau') \bigr) ,
\label{ansatz2}
\end{align} 
with initial conditions $Z_x=v/\pi K$, $v_\Lambda=v$ and $V_{\Lamb,ij}(u)=2\calD_{ij}\cos(2u)$. The form of $\Gamma_{1,k}$ and $\Gamma_{2,k}$ is strongly constrained by the statistical tilt symmetry (STS) due to the invariance of the disorder part of the action~(\ref{action}) in the time-independent shift $\varphi_{ia}(X)\to \varphi'_{ia}(X)=\varphi_{ia}(X)+w(x)$ with $w(x)$ an arbitrary function of $x$.\cite{Dupuis20} The STS yields
\begin{align}
\Gamma_k[\phipset] ={}& \Gamma_k[\phiset] + n \beta Z_x \int_x (\dx w)^2 \nonumber \\ & + Z_x \sum_{i,a} \int_X (\dx w)(\dx \phi_{ia}) . 
\end{align} 
This implies that $Z_x$ remains equal to its initial value and no other space derivative terms are allowed; for instance the term $(\dx\phi_{1a})(\dx\phi_{2a})$ is not possible. The term $(\dtau\phi_{1a})(\dtau\phi_{2a})$ is {\it a priori} not excluded by the STS but is not generated by the flow equation. Since the two copies are equivalent ($\calD_{11}=\calD_{22}$), the velocity $v_k$ is copy independent. In addition to $v_k$ one may define a $k$-dependent Luttinger parameter by $Z_x=v_k/\pi K_k$. The STS also ensures that the two-replica potential $V_{ij,k}(\phi_{ia},\phi_{jb})$ is a function of $\phi_{ia}-\phi_{jb}$ only. 

Thus the main quantities of interest are $K_k$, $v_k$ and the two-replica potential $V_{ij,k}(u)$. It is convenient to introduce the dimensionless function 
\beq 
\delta_{ij,k}(u) = - \frac{K^2}{v^2} \frac{V_{ij,k}''(u)}{k^3} .
\eeq 
For a single copy, the BG fixed point is characterized by a vanishing of $K_k$ and $v_k$: $K_k,v_k\sim k^\theta$ for $k\to 0$. The vanishing of $K_k$ implies that quantum fluctuations are suppressed at low energies and therefore a pinning of the field $\varphi(x,\tau)$ by the random potential. On the other hand the $\pi$-periodic function $\delta^*(u)=\lim_{k\to 0}\delta_k(u)$ exhibits cusps at $u=p\pi$ ($p$ integer). This cuspy nonanalytic form is related to the existence of metastable states.\cite{Balents96} At nonzero momentum scale, quantum tunneling between the ground state and these metastable states leads to a rounding of the nonanalyticity into a QBL. The latter is responsible for the vanishing of the optical conductivity $\sigma(\w)\sim \w^2$ in the low-frequency limit.\cite{Dupuis19,Dupuis20} 

With the ansatz~(\ref{ansatz1},\ref{ansatz2}), the disconnected propagator of the two-copy system is given by  
\beq
G_{d,ij,k}(q,i\w) = \beta \delta_{\w,0} \frac{v^2k^3}{K^2} \frac{\delta_{ij,k}(0)}{[Z_x q^2 + R_k(q,0)]^2} .
\eeq
{\it Stricto sensu} this expression is valid only for $|q|\ll k$ since the ansatz~(\ref{ansatz1},\ref{ansatz2}) is based on a derivative expansion. However, we expect $q$ to act as an infrared cutoff in the flow so that $G_{d,ij,k=0}(q,0)$ can be approximately obtained by setting $k\sim |q|$, i.e.,
\beq 
G_{d,ij,k=0}(q,i\w) \sim \beta \delta_{\w,0} \frac{\pi^2 \delta_{ij,|q|}(0)}{|q|} .
\eeq  
Since the intracopy correlation function $C_{ii}(X)$ is not modified by the inter-copy statistical correlations, we have $C_{ii}(X) \simeq 2\pi \delta^*(0) \ln|x|$.\cite{Dupuis20} The system exhibits chaos if $\lim_{x\to\infty}C_{12}(X)=0$ for any $\eps>0$, which requires $\lim_{k\to 0}\delta_{12,k}(0)=0$: The two copies are then statistically independent at long distances regardless of the (nonzero) difference in the random potentials $\xi_1(x)$ and $\xi_2(x)$. 
In the following we shall therefore consider the flow of 
\beq 
F_k = \frac{1+\eps^2}{1-\eps^2} \frac{\delta_{12,k}(0)}{\half[\delta_{11,k}(0)+\delta_{22,k}(0)]} ,
\label{Fkdef} 
\eeq 
with initial condition $F_\Lambda=1$.

\subsection{Flow equations} 

A detailed derivation of the flow equations for a single copy can be found in Ref.~\onlinecite{Dupuis20}. The generalization to two copies is straightforward and gives 
\begin{align} 
\dt\delta_{ii,k}(u) ={}& -3 \delta_{ii,k}(u) - l_1 K_k \delta''_{ii,k}(u) \nonumber \\ & \hspace{-1.5cm}
+ \pi \bar l_2 \{ \delta_{ii,k}''(u) [\delta_{ii,k}(u)-\delta_{ii,k}(0)] + \delta'_{ii,k}(u)^2 \} ,  \label{rgeq1} \\ 
\dt\delta_{12,k}(u) ={}& -3 \delta_{12,k}(u) - l_1 K_k \delta''_{12,k}(u) \nonumber \\ &  \hspace{-1.5cm}
+ \pi \bar l_2 \{ \delta_{12,k}''(u) [\delta_{12,k}(u)-\delta_{ii,k}(0)] + \delta'_{12,k}(u)^2 \} , \label{rgeq2} 
\end{align}
and 
\begin{equation} 
\dt K_k = \theta_k K_k , \qquad
\dt (K_k/v_k) = 0 ,
\end{equation}
where
  
\begin{equation}
    \theta_k = z_k-1 = \frac{\pi}{2} \delta''_{ii,k}(0) \bar m_\tau 
\label{thetak}
\end{equation}
with $z_k$ being the running dynamical critical exponent.
The thresholds functions $l_1$, $\bar l_2$ and $\bar m_\tau$ are defined in Ref.~\onlinecite{Dupuis20}. Note that the $\pi$ periodicity of $\delta_{ij,k}(u)$, as well as the property $\int_0^\pi du\, \delta_{ij,k}(u)=0$, are maintained by the flow equations.

\section{Disorder chaos} 
\label{sec_dchaos} 

We consider only the case where the parameters of the microscopic action (i.e., the initial conditions of the RG flow) are such that the system is in the BG phase. The flow equations are integrated numerically using the fourth-order Runge-Kutta method with adaptative step size. The functions 
\beq
\delta_{ij,k}(u) = \sum_{n=1}^{\nmax} \delta_{ij,n,k} \cos(2nu) 
\label{harmonics}
\eeq 
are expanded in circular harmonics with $\nmax$ in the range $[100-800]$. Note that $\delta_{ij,0,k}$ necessary vanishes since $\int_0^\pi du\, \delta_{ij,k}(u)=0$. 

\subsection{Approach to the BG fixed point} 
\label{subsec_approach}

As expected, the RG equations for $\delta_{ii,k}(u)$, $K_k$ and $v_k$ are identical to the one-copy case.\cite{Dupuis20} The function $\delta_{ii,k}(u)$ approaches the $\pi$-periodic fixed-point solution 
\beq
\delta^*(u) = \frac{1}{2\pi\bar l_2} \left[ \left( u - \frac{\pi}{2} \right)^2 - \frac{\pi^2}{12} \right] , \quad u \in [0,\pi] ,
\label{deltafpsol}
\eeq
which exhibits cusps at $u=p\pi$ ($p$ integer). The BG fixed point is a ``critical'', scale-invariant, fixed point as far as the disorder correlator $\delta_{ii,k}(u)$ is concerned, i.e., in the zero-frequency sector. The finite localization length, which characterizes the BG phase, appears only in the nonzero-frequency sector of the theory, a feature which is related to the nonanalytic structure of the propagator at zero frequency.\cite{Dupuis20} 

\subsubsection{Quantum boundary layer} 
\label{subsubsec_QBL}

For any nonzero momentum scale $k$, the cusp singularity at $u=p\pi$ is rounded into a boundary layer. In the vicinity of the BG fixed point, $K_k\to 0$, the solution can be written in the form 
\beq 
\delta_{ii,k}(u) = \delta_{ii,k}(0) + K_k f\left( \frac{u}{K_k} \right) 
\label{deltaQBL}
\eeq
near $u=0$ and for an arbitrary value of the ratio $u/K_k$. The $k$-independent even function satisfies $f(0)=f'(0)=0$ and $f''(0)<0$. From~(\ref{rgeq1}) we obtain 
\beq
\dt \delta_{ii,k}(0) \simeq -3 \delta_{ii,k}(0) -l_1 f'' + \pi \bar l_2(f'' f + f'{}^2 ) 
\label{deltaQBL1}
\eeq 
using $K_k\to 0$ and $\dt K_k=\theta_k K_k\to 0$. The right-hand side must be independent of $x=u/K_k$ and equal to $-3\delta_{ii,k}(0)-l_1f''(0)$ since $f(0)=f'(0)=0$, i.e., 
\beq
-l_1 f'' + \pi \bar l_2(f''f+f'{}^2) = -l_1 f''(0) . 
\label{deltaQBL2}
\eeq
This yields 
\beq 
f(x) = \frac{l_1}{\pi \bar l_2} \llbrace 1 - \left[1 - \frac{\pi\bar l_2}{l_1} f''(0) x^2 \right]^{1/2} \rrbrace . 
\label{deltaQBL3}
\eeq 
Since the solution~(\ref{deltaQBL}) must approach the fixed-point solution $\delta^*(u)$ when $K_k\to 0$, we deduce $f''(0)=-\pi/4l_1\bar l_2$. 

From~(\ref{deltaQBL1}) and (\ref{deltaQBL2}) we also obtain 
\beq 
\dt \delta_{ii}(0) = - 3 \delta_{k,ii}(0) - l_1 f''(0) , 
\eeq 
i.e., 
\beq 
\delta_{ii,k}(0) = C e^{-3t} - \frac{l_1}{3} f''(0) . 
\eeq 
Since $\delta_{11,k}(0)$ approaches a finite value as $t\to -\infty$, the constant $C$ must necessarily vanish. From the flow equation~(\ref{rgeq1}) it is easy to see that the relevant eigenvalue 3 is associated with a constant solution $\delta_k(u)=\const$, which is not allowed as it violates the condition $\int_0^\pi du\, \delta_{ii,k}(u)=0$. We therefore obtain 
\beq
\delta_{ii,k}(0) = - \frac{l_1}{3}f''(0) = \frac{\pi}{12\bar l_2} = \delta^*(0) ,
\eeq 
which is the expected result in the limit $k\to 0$.  For $k>0$, we expect
\beq 
\delta_{ii,k}(0) = \delta^*(0) - \frac{l_1}{\pi\bar l_2} K_k  
\label{deltaQBL4}
\eeq 
to leading order in $K_k$,\cite{not3} where the prefactor of $K_k$ is determined by requiring that $\delta_{ii,k}(u)-\delta^*(u)$ vanishes to order $K_k$ when $|u|/K_k\gg 1$. The QBL  defined by~(\ref{deltaQBL},\ref{deltaQBL3},\ref{deltaQBL4}) is in very good agreement with the numerical solution of the flow equations (see the discussion of the QBL and CBL in Sec.~\ref{subsubsec_CBL} and Fig.~\ref{fig_BL}).

The preceding results imply the deviation from the BG fixed point
\begin{align}
g_k(u) &= \delta_{ii,k}(u)-\delta^*(u) \nonumber\\ &
\simeq 
\llbrace \begin{array}{lll} 
	- \frac{l_1}{\pi\bar l_2} K_k + \frac{|u|}{2\bar l_2} & \mbox{if} & |u|\ll K_k , \\
	0 & \mbox{if} & |u|\gg K_k . 
	\end{array} 
\right. 
\end{align}
We conclude that when $K_k\to 0$ (i.e., $k\to 0$) the function $g_k(u)$, which characterizes the approach to the BG fixed point {\it via} the QBL, tends to $\sim K_k\sh^{(K)}_{u,0}$ where $\sh^{(K)}_{u,0}$ is the $\pi$-periodic Kronecker comb.\cite{not8} In the next section we shall see how this result can be reproduced from a linear analysis of the perturbations about the fixed point.

\subsubsection{Linear analysis} 
\label{subsubsec_linearanalysis_QBL}

Let us consider a small perturbation about the BG fixed point $(K^*=0,\delta^*(u))$,
\beq 
\begin{split}
\delta_{ii,k}(u) &= \delta^*(u) + g(u) e^{-\lamb t} , \\ 
K_k &= K^* + K e^{-\lamb t} , 
\end{split}
\eeq 
where $g(u)$ and $K$ (not to be confused with the bare value $K_{\Lambda}$ of the Luttinger parameter) are $k$ independent. To first order in $g$ and $K$, we obtain the flow equations 
\begin{align}
\lamb g(u) ={}& 3 g(u) + K l_1 \delta^*{}''(u) - \pi\bar l_2 \{ g''(u) [\delta^*(u)-\delta^*(0)] \nonumber \\ & + \delta^*{}''(u) [g(u)-g(0)]  + 2 \delta^*{}'(u) g'(u) \}
\label{linear1}
\end{align} 
and 
\beq 
\lamb K = -\theta K ,
\label{linear2}
\eeq
with $\theta=\lim_{k\to 0}\theta_k$. We now expand the functions $g(u)$ and $\delta^*(u)$ in circular harmonics as in~(\ref{harmonics}) with 
\beq  
\delta^*_n = \frac{1}{2\pi\bar l_2 n^2} .
\eeq 
Note that this expression implies that 
\beq 
\delta^*{}''(u) = \frac{1}{\pi\bar l_2}[1-\pi\sh(u)] ,
\label{deltafppp} 
\eeq 
where $\sh(u)$ the $\pi$-periodic Dirac comb,\cite{not8} and differs from the naive result $\delta^*{}''(u) =1/\pi\bar l_2$ (which violates the condition $\int_0^\pi du\, \delta^*{}''(u)=0$) obtained from~(\ref{deltafpsol}). We thus rewrite Eqs.~(\ref{linear1}) and (\ref{linear2}) as
\beq 
\begin{split} 
\lamb g_n &= \sum_{n'=1}^{\nmax} A_{n,n'} g_{n'} - \frac{2l_1}{\pi\bar l_2} K , \\ 
\lamb K &= -\theta K ,
\end{split} 
\label{linear3}
\eeq 
with 
\begin{align}
A_{n,n'} ={}& \left( 3 - \frac{\pi^2}{3}n^2 \right) \delta^{(K)}_{n,n'} \nonumber \\ & 
- 2 + n^2 \left[ \frac{1}{(n+n')^2} + \frac{1-\delta^{(K)}_{n,n'}}{(n-n')^2} \right] ,
\label{Adef}    
\end{align}
where $\delta^{(K)}_{n,n'}$ denotes the Kronecker delta. 

The solution with eigenvalue $\lambda=-\theta$, corresponding to a vanishing of the Luttinger parameter $K_k\sim e^{\theta t}$, can be found analytically in the limit $\nmax\to \infty$. Setting $K=1$ and $g_n=-l_1/\pi \bar l_2 \nmax$, and using 
\beq 
\sum_{n'=1}^{\nmax} (A_{n,n'}+2) = 2 + \calO\left(\frac{1}{\nmax}\right) ,
\label{Aprop} 
\eeq 
one easily sees that the first equation in~(\ref{linear3}) is satisfied to leading order in $1/\nmax$. Thus, for $\nmax\to\infty$, we obtain the linear perturbation 
\beq 
K=1, \qquad g(u)= - \frac{l_1}{\pi \bar l_2}\sh^{(K)}_{u,0} .
\label{gnamxinf} 
\eeq 
The result $g(u)\propto \sh^{(K)}_{u,0}$ agrees with the QBL analysis of Sec.~\ref{subsubsec_QBL}.

\begin{figure}
    \centerline{\includegraphics[scale=1]{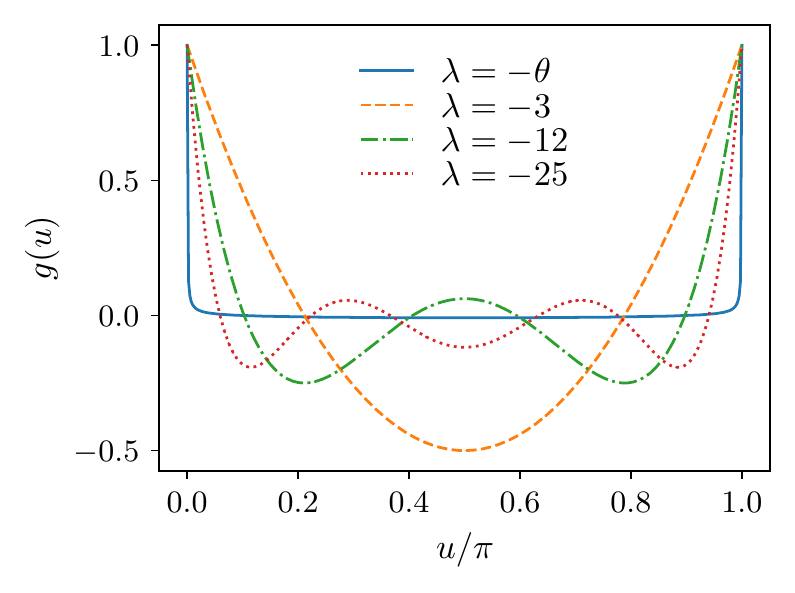}}
\caption{Solutions $g(u)=\sum_{n=1}^{\nmax} g_n \cos(2nu)$ deduced from the numerical solution of Eqs.~(\ref{linear3}) with $\nmax=400$ and $\theta=0.5$, and corresponding to the largest eigenvalues: $\lambda=-\theta=-0.5$ (solid line) and $\lambda\simeq -3.00/-12.00/-25.00$ (dashed and dotted lines). The functions are normalized by the condition $g(0)=1$.}
\label{fig_linearQBL} 
\end{figure}

This result is confirmed by the numerical solution of the linear system~(\ref{linear3}). As $\nmax$ increases, we find a solution $g(u)$, associated with an eigenvalue $\lambda\simeq-\theta$, and which becomes increasingly localized about $u=p\pi$ ($p$ integer) with the ratio $g(p\pi)/K$ taking a nonzero limit, in agreement with~(\ref{gnamxinf}). There are also $\nmax$ eigenvectors with $K=0$ and a function $g(u)$ which typically extends over the whole interval $[0,\pi]$ (see Fig.~\ref{fig_linearQBL}). The largest eigenvalues converge to $\{-3,-12,-25,-42,\cdots\}$ when $\nmax\to\infty$. The convergence is fast and, for the largest eigenvalues, already obtained with a two-digit precision for $\nmax=400$. 
Although $\theta$ is not precisely known,\cite{Dupuis20} it satisfies $\theta<3$; $-\theta$ is therefore the largest eigenvalue and controls the approach to the BG fixed point.  

If we set $K\equiv K_\Lambda=0$ in the flow equation~(\ref{rgeq1}) the cusp in $\delta_{ii,k}(u)$ arises for $k>0$.\cite{Dupuis20} This finite-scale singularity is not accounted for in the linear analysis. Indeed, if we set $K=0$ in~(\ref{linear1}) and (\ref{linear2}), we find that the function $g(u)e^{\lambda t}$ goes smoothly to zero and the fixed point $\delta^*(u)$ is recovered only for $k=0$. Thus it seems that the boundary layer induced by a nonzero $K_k$ is a necessary condition for the linear analysis to be valid.

\subsection{Escape from the BG fixed point} 
\label{subsec_escape} 

\subsubsection{Chaotic behavior and overlap length}

\begin{figure}
	\centerline{\includegraphics[scale=1]{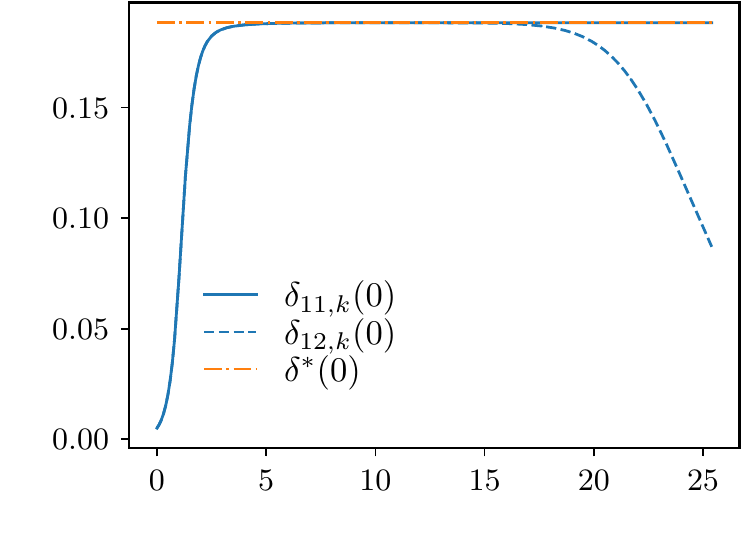}}
	\vspace{-0.25cm}
	\centerline{\includegraphics[scale=1]{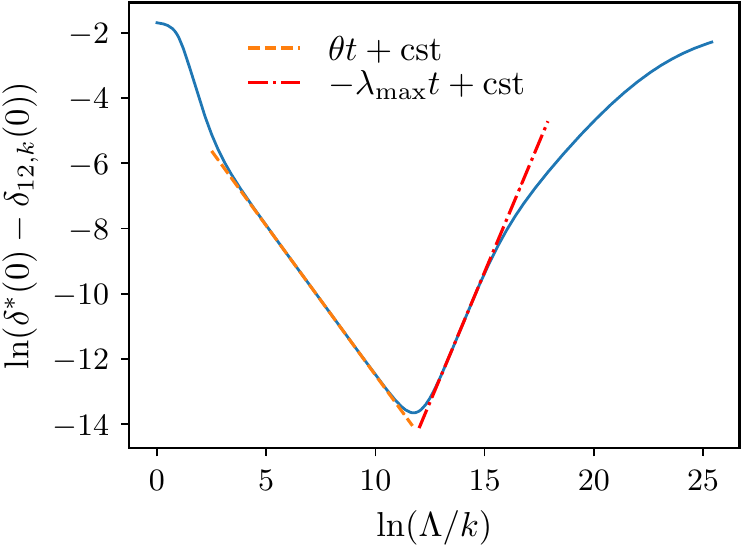}}
	\caption{(top) $\delta_{11,k}(0)=\delta_{22,k}(0)$ and $\delta_{12,k}(0)$ vs $\ln(\Lambda/k)=-t$ as obtained from the numerical solution of the flow equations with $\nmax=800$ and $\eps\simeq 3\times10^{-8}$. When $\eps\neq 0$, $\delta_{12,k}(0)$ first approaches the BG fixed-point value $\delta^*(0)$ but is eventually suppressed, $\lim_{k\to 0}\delta_{12,k}(0)=0$, thus showing that the two copies become statistically uncorrelated at long distances. (bottom) The ln-ln plot shows that the approach of $\delta_{12,k}(0)$ to its fixed point value $\delta^*(0)$ is controlled by the exponent $-\theta$ and the escape by $\lambda_{\rm max}\simeq 1.6$. (These figures are obtained for $K=0.1$, other figures use $K=0.4$.)}
	\label{fig_delta12}
\end{figure}

When $\eps=0$ the two copies are identical, $\delta_{12,k}(u)=\delta_{ii,k}(u)$, and $\delta_{12,k}(u)$ approaches $\delta^*(u)$ when $k\to 0$. For small but nonzero $\eps$, implying $\delta_{12,\Lamb}(u)\neq\delta_{ii,\Lamb}(u)$, $\delta_{k,12}(u)$ is first attracted to the BG fixed-point solution $\delta^*(u)$ but is eventually suppressed as shown in Fig.~\ref{fig_delta12}: 
\beq 
\lim_{k\to 0}\delta_{12,k}(u) = 0 .
\eeq 
Linearizing the equation $\dt\delta_{12,k}$, we find 
\beq 
\delta_{12,k}(u)\sim k^{-3+\pi^2/3}\cos(2u) \quad \mbox{for} \quad k\to 0
\label{delta12_sup}
\eeq 
(higher-order harmonics decay faster), which gives $G_{d,12,k=0}(q,0)\sim |q|^{-4+\pi^2/3}$, so that   
\beq 
C_{12}(X) \sim \frac{1}{|x|^{-3+\pi^2/3}}
\eeq   
decays with the exponent $\pi^2/3-3\simeq 0.2899$. We conclude that the BG phase exhibits chaos. 

\begin{figure}
	\centerline{\includegraphics[scale=1]{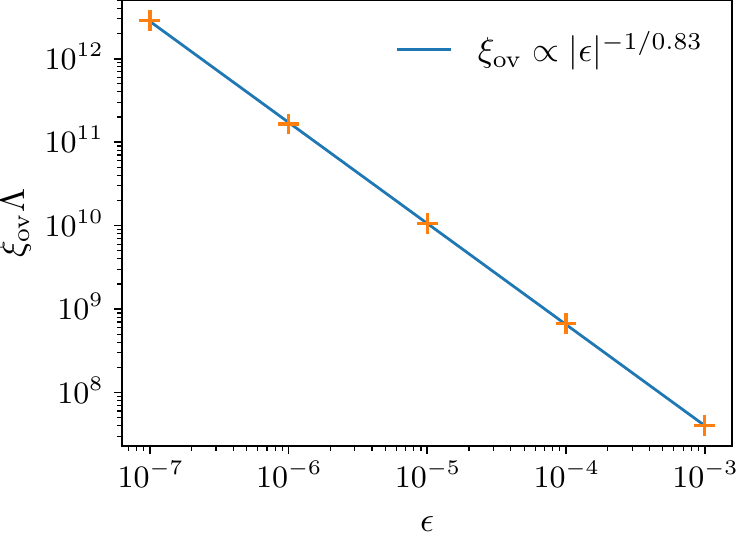}} 
	\caption{$\xiov$ vs $\eps$ in a log-log plot as obtained from the criterion $F_{k=1/\xiov}=0.1$. The blue line shows a linear fit corresponding to the power-law behavior~(\ref{xiov}) with $\alpha\simeq0.83$.}
	\label{fig_xiov} 
\end{figure}

\begin{figure}[b]
	\centerline{\includegraphics[width=7.5cm]{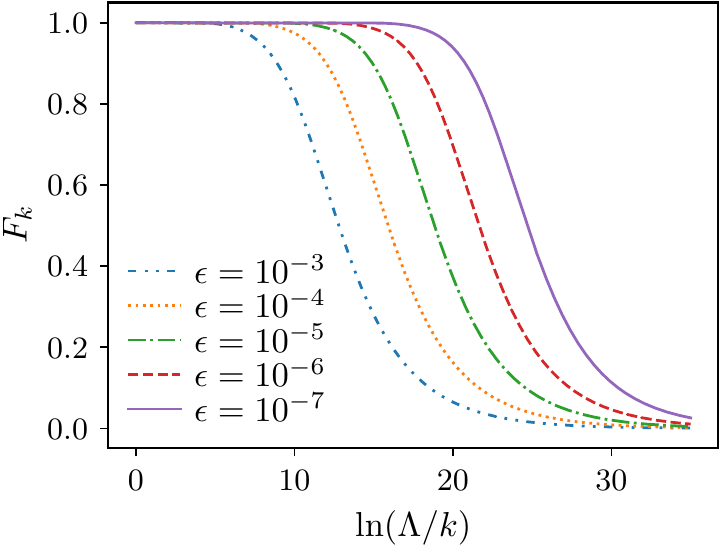}}
	\vspace{0.2cm}
	\centerline{\includegraphics[width=7.5cm]{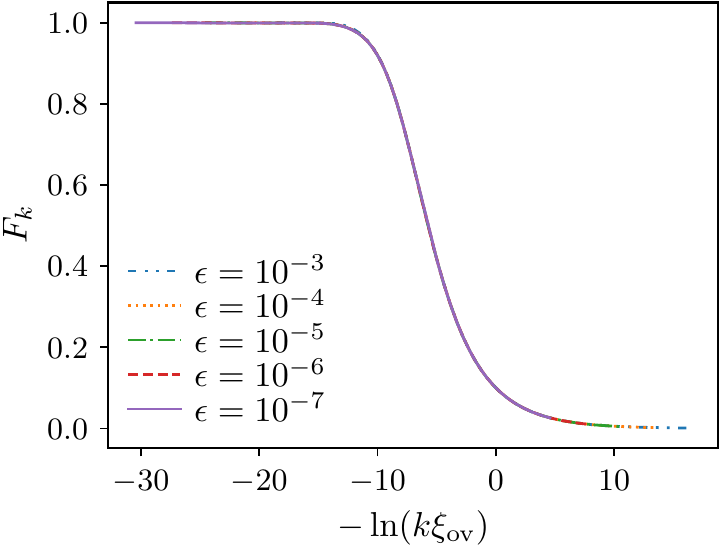}}
	\caption{(top) $F_k$ vs $k$ for various values of $\eps$. Statistical correlations between the two copies are lost when $1/k$ is larger than the overlap $\xiov$ length defined by $F_{k=1/\xiov}=0.1$. (bottom) $F_k$ vs $k\xiov$ showing the data collapse expected from the scaling form~(\ref{scaling}).}  
	\label{fig_scaling} 
\end{figure}

We can define a characteristic (overlap) length $\xiov\equiv\xiov(\eps)$ associated with the instability of the BG fixed point when $\eps\neq 0$ and signaling the loss of statistical correlations between the two copies at large length scales. We use the criterion $F_{k=1/\xiov}=\gamma$ where $\gamma\ll 1$ is an arbitrary number and $F_k$ is defined by~(\ref{Fkdef}). $\xiov$ diverges for $\eps\to 0$ as a power law (Fig.~\ref{fig_xiov}), 
\beq 
\xiov \propto |\eps|^{-1/\alpha} ,
\label{xiov} 
\eeq 
where $\alpha$ is the chaos exponent. If we plot $F_k$ as a function of $k\xiov$ for various values of $\eps$, we observe a data collapse thus showing that $F_k$ satisfies the one-parameter scaling form 
\beq 
F_k = \calF(k\xiov) ,
\label{scaling} 
\eeq 
where $\calF(x)$ is a universal scaling function, as expected for a scale-invariant fixed point with a single relevant direction (Fig.~\ref{fig_scaling}). 

On can also obtain the chaos exponent directly from the flow equations. For $\eps\to 0$, when $\delta_{12,k}(0)$ is near its fixed point value $\delta^*(0)$, one has 
\beq 
\delta_{12,k}(0) \simeq \delta^*(0) + A e^{\theta t} + B e^{-\lambda_{\rm max}t} ,
\label{alphadef2} 
\eeq 
as shown in Fig.~\ref{fig_delta12}. The leading irrelevant eigenvalue $-\theta$, as discussed in  Sec.~\ref{subsubsec_linearanalysis_QBL}, controls the approach to the BG fixed point. The relevant eigenvalue $\lambda_{\rm max}$ controls the departure from the fixed point at very long RG time $|t|$. $\lambda_{\rm max}$ also determines the divergence of the overlap length when $\eps\to 0$, 
\beq
\xiov \sim |\delta_{12,\Lambda}(0) - \delta_{ii,\Lambda}(0)|^{1/\lambda_{\rm max}}\sim |\eps|^{-2/\lambda_{\rm max}} , 
\eeq 
so that $\alpha=\lambda_{\rm max}/2$. The estimate of the chaos exponent obtained from~(\ref{alphadef2}) is in very good agreement with the calculation of $\xiov$ using the criterion $F_{k=1/\xiov}=\gamma$. The results are shown in Table~\ref{table_alpha} for various values of $\nmax$. 

\begin{table*}
	\caption{Chaos exponent $\alpha$ vs number $\nmax$ of harmonics used in the numerics obtained from $\xiov$,  $\delta_{12,k}(0)-\delta^*(0)$ or the linear analysis [Eqs.~(\ref{CBL3})].} 
	\begin{tabular}{lccccccccccccc}
		\hline \hline
		$\nmax$ & 100 & 200 & 300 & 400 & 500 & 600 & 1000 & 10000 & 20000 & 30000& 40000& 50000& 100000\\
		from $\xiov$ &0.803 &0.816 &0.826 &0.831 & 0.832& &&&&&&&\\
		from $\delta_{12,k}(0)-\delta^*(0)$ &0.800&0.811&0.817&0.820& 0.823& & & & &&&& \\  
		from linear analysis &0.717&0.748&0.764&0.773&0.781&0.786&0.800&0.847&0.857&0.862&0.865&0.868&0.876 \\ 
		\hline
	\end{tabular}
	\label{table_alpha}
\end{table*}

\subsubsection{Chaos boundary layer} 
\label{subsubsec_CBL} 

\begin{figure}[t]
    \centerline{\includegraphics[scale=1]{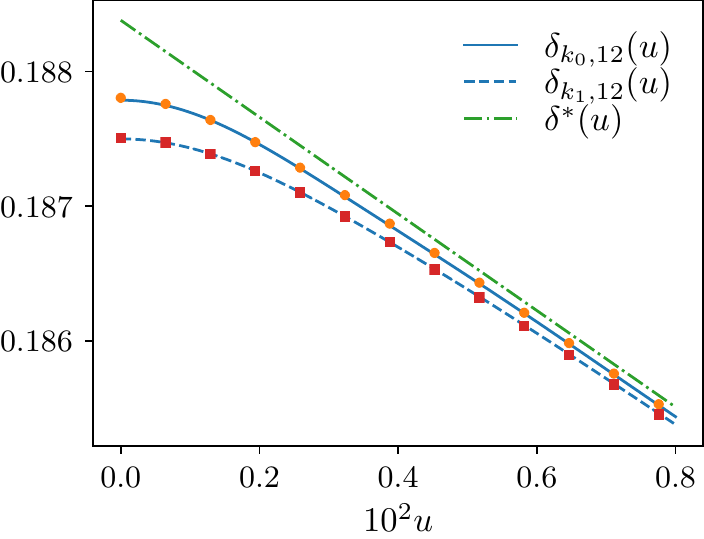}}
    \caption{$\delta_{12,k}(u)$ near $u=0$ as obtained from the numerical solution of the flow equations (lines) and the analytic expression~(\ref{deltaCBL3}) (symbols).  $\delta_{12,k_0=\Lambda e^{-4.5}}$ shows the QBL that forms during the approach to the BG fixed point whereas $\delta_{12,k_1=\Lambda e^{-18.5}}$ shows the CBL due to the escape of the fixed point. When using expression~(\ref{deltaCBL3}), the value of $K_{{\rm tot},k}$ is obtained from the numerical solution of the flow equations.} 
    \label{fig_BL} 
\end{figure}

The boundary layer analysis of Sec.~\ref{subsubsec_QBL} can be generalized to the case where $\eps\neq 0$. Since the flow equation of $\delta_{ii,k}(u)$ and $K_k$ are independent of $\eps$, the intracopy disorder correlator $\delta_{ii,k}(u)$ is still given by~(\ref{deltaQBL}) when $|u|,K_k\ll 1$. The equation for $\delta_{12,k}(u)$ can be written as
\begin{align}
\dt\delta_{12,k}(u) ={}& -3 \delta_{12,k}(u) - l_1 \left( K_k + \frac{\pi \bar l_2}{l_1} \hat K_k \right) \delta''_{12,k}(u) \nonumber \\ &  \hspace{-1.5cm}
+ \pi \bar l_2 \{ \delta_{12,k}''(u) [\delta_{12,k}(u)-\delta_{12,k}(0)] + \delta'_{12,k}(u)^2 \} , 
\label{deltaCBL1}
\end{align}
where 
\beq 
\hat K_k = \delta_{ii,k}(0) - \delta_{k,12}(0) .
\label{deltaCBL2} 
\eeq 
We assume that the system is near the BG fixed point, so that both $K_k,\hat K_k$ and their derivatives $\dt K_k,\dt \hat K_k$ are small.\cite{not5} Near $u=0$, but for an arbitrary ratio $u/K_{{\rm tot},k}$, the solution of~(\ref{deltaCBL1},\ref{deltaCBL2}) can be written in the form 
\beq 
\delta_{12,k}(u) = \delta_{12,k}(0) + K_{{\rm tot},k}  f \left( \frac{u}{K_{{\rm tot},k}} \right) , 
\label{deltaCBL3} 
\eeq 
where 
\beq 
K_{{\rm tot},k} = K_k + \frac{\pi\bar l_2}{l_1} \hat K_k . 
\eeq 
Using the fact that $K_{{\rm tot},k}$ and $\dt K_{{\rm tot},k}$ are small, Eq.~(\ref{deltaCBL1}) implies that $f(x)$ satisfies~(\ref{deltaQBL2}) and is therefore given by~(\ref{deltaQBL3}) with $f''(0)=-\pi/4l_1\bar l_2$. This is expected since when $\eps=0$, one has $K_{{\rm tot},k}=K_k$ and $\delta_{12,k}(u)=\delta_{ii,k}(u)$ is given by~(\ref{deltaQBL}). Similarly to~(\ref{deltaQBL4}) we find 
\beq 
\delta_{12,k}(0) = \delta^*(0) - \frac{l_1}{\pi\bar l_2} K_{{\rm tot},k} .
\label{deltaCBL4} 
\eeq
When $\delta_{12,k}(u)$ approaches the fixed point, i.e., when $K_{{\rm tot},k}\simeq K_k$, the decreasing width of the boundary layer is controlled by quantum fluctuations as discussed in Sec.~\ref{subsubsec_QBL} ($\delta_{12,k}(u)\simeq \delta_{ii,k}(u)$ in that case). At latter RG times $|t|$, when $K_{{\rm tot},k}\simeq (\pi\bar l_2/l_1)\hat K_k$, the width of the boundary layer increases as a result of the loss of statistical correlations between the two copies due to the chaotic behavior of the system. Equations~(\ref{deltaCBL3},\ref{deltaCBL4}) are in very good agreement with the numerical solution of the flow equations as shown in Fig.~\ref{fig_BL}.

\subsubsection{Linear analysis} 
\label{subsubsec_CBL_linear}

We now consider a linear analysis of the perturbations about the BG fixed point in the case $\eps\neq 0$. Writing 
\beq
\begin{split} 
\delta_{ii,k}(u) &= \delta^*(u) + g_{ii}(u) e^{-\lamb t} , \\
\delta_{12,k}(u) &= \delta^*(u) + g_{12}(u) e^{-\lamb t} , \\
K_k &= K^* + K e^{-\lamb t} 
\end{split}
\eeq 
(note that $\hat K_k$, introduced in the preceding section, is not an independent variable), we obtain 
\begin{align}
\lamb g_{12}(u) ={}& 3 g_{12}(u) + K l_1 \delta^{*}{}''(u) - \pi\bar l_2 \{ g''_{12}(u) [\delta^*(u)-\delta^*(0)] \nonumber \\ & + \delta^*{}''(u) g_{12}(u)  + 2 \delta^*{}'(u) g_{12}'(u) \}
\label{linear4}
\end{align} 
to first order in $g_{12}$ while $g_{ii}$ and $K$ satisfy~(\ref{linear1},\ref{linear2}). Expanding both $g_{ij}(u)$ and $\delta^*(u)$ in circular harmonics yields the linear system
\beq 
\begin{split} 
	\lamb g_{ii,n} &= \sum_{n'=1}^{\infty} A_{n,n'} g_{ii,n'} - \frac{2l_1}{\pi\bar l_2} K , \\ 
	\lambda g_{12,n} &= \sum_{n'=1}^\infty B_{n,n'} g_{12,n'} - \frac{2l_1}{\pi\bar l_2} K  , \\
	\lamb K &= -\theta K ,
\end{split} 
\label{CBL3} 
\eeq 
where $A_{n,n'}$ is defined in~(\ref{Adef}) and 
\begin{align}
B_{n,n'} = A_{n,n'} + 2 .   
\label{Bdef} 
\end{align}

Using 
\beq 
\sum_{n'=1}^{\infty} B_{n,n'} = 2 , 
\eeq 
which follows from~(\ref{Aprop}), we see that $K=0$ and $g_{12,n}=\const$, i.e., 
\beq 
g_{12}(u) = \pi \sh(u)-1 ,
\label{sol2}
\eeq
is solution with eigenvalue $\lambda=2$ and satisfies $\int_0^\pi du\, g_{12}(u)=0$.  It qualitatively reproduces the result obtained from the boundary layer analysis in Sec.~\ref{subsubsec_CBL} although the latter gives a Kronecker comb and not a Dirac comb (see the discussion in Sec.~\ref{subsubsec_linearanalysis_QBL}). We do not expect the difference between the Kronecker and Dirac combs to bear a particular physical meaning. In both cases the singular function $g_{12}(u)$ originates from the boundary layer near $u=p\pi$, be it a QBL or a CBL [Eq.~(\ref{deltaCBL3})]. Similarly we find that $g_{12}(u)$ defined by~(\ref{sol2}), together with
\beq 
K = \frac{\pi\bar l_2}{l_1} (\theta+2) , 
\eeq 
is solution with eigenvalue $\lambda=-\theta$. 

Let us now look for the other eigenfunctions, with $\lambda\neq 2,-\theta$ and therefore $K=0$, in the form 
\beq 
g(u) = \pi \sh(u)-1 + h(u) , 
\eeq
where $h(u)$ is assumed to be free of Dirac peaks but its derivative may be discontinuous at $u=p\pi$. From the equations satisfied by $g(u)$ and $\pi \sh(u)-1$, we easily obtained 
\begin{multline}
(\lambda-2)[\pi \sh(u)-1] + \lambda h(u) = 3 h(u) - \pi\bar l_2 \{ 
\delta^*{}''(u) h(u) \\  + h''(u) [\delta^*(u)-\delta^*(0)]  + 2 \delta^*{}'(u) h'(u) \} .
\label{linear5}
\end{multline} 
Collecting all terms involving Dirac peaks,\cite{not1} we obtain 
\begin{align}
(\lambda-2) \pi \sh(u) &= \pi \sh(u) h(u) \nonumber \\
&= \pi \sh(u) h(0) .
\end{align}
This equation is satisfied if 
\beq
h(0) = \lambda - 2 . 
\label{cond}
\eeq 
The terms free of Dirac peaks lead to the equation 
\beq
0 = (4-2\lambda)(h-1) + u(\pi-u) h'' +2(\pi-2u) h' 
\eeq
for $u\in [0,\pi]$. Setting $h=1+f$ and introducing $x=u/\pi$, we finally obtain  
\beq
0 = (4-2\lambda)f + x(1-x) f'' +2(1-2x) f' ,
\label{linear6}
\eeq 
where the function $f(x)$ must satisfy
\beq
\int_0^1 dx \, f(x) = -1 . 
\label{linear7}
\eeq 
Equation~(\ref{linear6}) was studied by DLD.\cite{Duemmer07} The solutions that are symmetric about $x=1/2$ in the interval $[0,1]$ (this condition follows from $h(u)$ being even and $\pi$ periodic) can be expressed in terms of hypergeometric functions. The condition~(\ref{linear7}) of integrability selects a discrete set of values of $\lambda$, for which the hypergeometric function becomes a polynomial function of finite order, 
\beq 
f(x) = \sum_{m=0}^{m_0} c_m x^m .
\eeq 
For $f(x)$ to be solution of~(\ref{linear6}), we must require 
\beq 
c_{m+1} = c_m \frac{m(m+3)-4+2\lambda}{(m+1)(m+2)} ,
\eeq
whereas $c_0$ is determined from~(\ref{linear7}). Imposing $c_{m_0+1}=0$ then gives 
\beq
\lambda= 2 - \frac{m_0(m_0+3)}{2} .
\label{spectrum} 
\eeq 
For $m_0=0$ we obtain $\lambda=2$ and $f(x)=-1$, i.e., $h(u)=0$, which reproduces the solution~(\ref{sol2}). The choice $m_0=1$, and more generally $m_0$ odd, must be discarded since the corresponding solutions do not satisfy $f(0)=f(1)$. For $m_0=2$, one finds $\lambda=-3$ and 
\beq 
f(x) = - 6 (1-5x + 5 x^2 ) . 
\eeq  
The condition~(\ref{cond}), $\lambda=2+h(0)=3+f(0)$, is satisfied. The next solution ($m_0=4$) corresponds to $\lambda=-12$ and 
\beq 
f(x)= -15 (1-14 x+56 x^2-84 x^3+42 x^4) , 
\eeq 
and satisfies~(\ref{cond}). All other solutions can be obtained similarly and are associated with eigenvalues that are more and more irrelevant as $m_0$ increases. The negative eigenvalue  spectrum $\{-3,-12,-25,-42,-63,-88,\cdots\}$ is the same as that obtained numerically for the approach to the BG fixed point. 

\begin{figure}
\centerline{\includegraphics[scale=1]{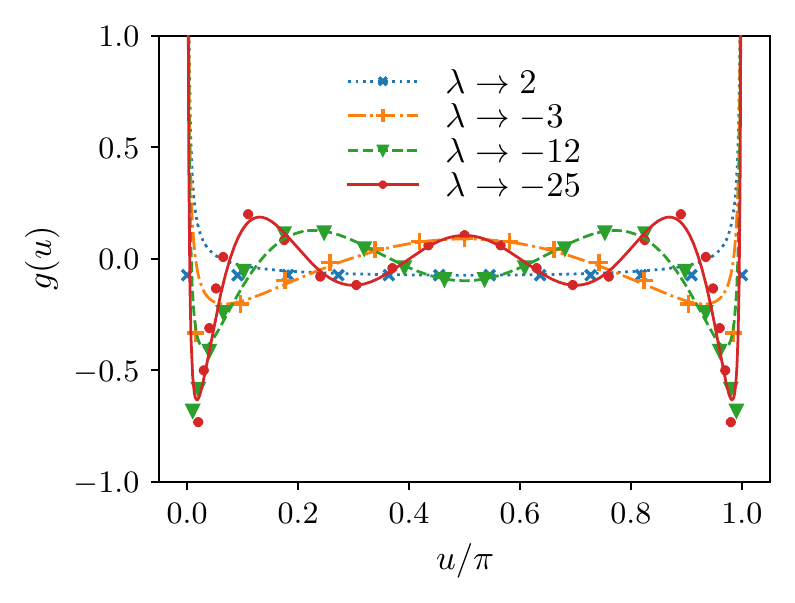}}
\caption{Functions $g(u)$, corresponding to the four largest eigenvalues, obtained from the numerical solution of~(\ref{CBL3}) with $\nmax=1000$ (symbols). Away from the points $u=0$ and $u=\pi$, these functions are well approximated by the functions $f(u)$ solutions of~(\ref{linear6}) associated with the eigenvalues $2/-3/-12/-25$, with an appropriate normalization (lines).}
\label{fig_linearCBL}
\end{figure}

\begin{figure}
\centerline{\includegraphics[scale=1]{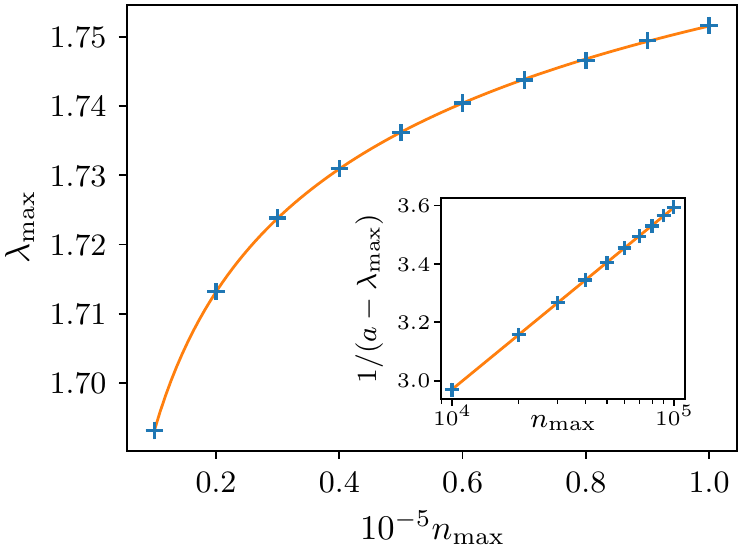}}
\caption{Largest eigenvalue $\lambda_{\rm max}$ vs $\nmax$ obtained from the numerical solution of~(\ref{CBL3}). The continuous (orange) line shows the logarithmic convergence~(\ref{lambvsnmax}) of $\lambda_{\rm max}$ with $\nmax$.}
\label{fig_convergence}
\end{figure}

In Fig.~\ref{fig_linearCBL} we show the solution $g(u)$ obtained from a numerical solution of~(\ref{CBL3}) with a finite number $\nmax$ of circular harmonics. We find that there is a single positive eigenvalue (in agreement with the analytic results), associated with a function $g(u)$ which, as $\nmax$ increases, is more strongly peaked near $u=p\pi$, with however a nonzero value away from these two points to ensure that $\int_0^\pi du\,g(u)=0$. This behavior is in qualitative agreement with~(\ref{sol2}). The functions $g(u)$ associated with negative eigenvalues are also strongly peaked near $u=0$ and $u=\pi$, and their behavior away from these two points is well approximated by the function $f(u)=g(u)-\pi\sh(u)$ found analytically above (Fig.~\ref{fig_linearCBL}). The convergence with $\nmax$ of the eigenvalues to the spectrum~(\ref{spectrum}) is however extremely slow. Even for a relatively large value $\nmax=100\,000$ we find that $\lambda_{\rm max}\simeq 1.75$ is still far from its expected converged value $\lambda_{\rm max}=2$. Our results agree with a logarithmic convergence, 
\beq
\lambda_{\rm max}(\nmax) \simeq a - \frac{b}{c+\ln(\nmax)} 
\label{lambvsnmax}
\eeq 
with $a\simeq 2.03$, $b\simeq 3.70$ and $c\simeq 1.77$, as shown in Fig.~\ref{fig_convergence}. 

The various values of the chaos exponent obtained numerically from either $\xiov$, $\delta_{12,k}(0)-\delta^*(0)$ or the linear analysis, are shown in Table~\ref{table_alpha}. Our analytic result $\lambda_{\rm max}=2$ implies a chaos exponent $\alpha=\lambda_{\rm max}/2=1$. 

The slow convergence of $\lambda_{\rm max}$ with $\nmax$ implies that it is necessary to probe the system at very long length scales to observe the value $\alpha=1$ of the chaos exponent. Any finite length $L$ indeed introduces an effective upper cutoff $\nmax\sim 1/L$ on the number of circular harmonics in the Fourier series expansion of the function $\delta_{12,k}(u)$. In particular this means that the critical behavior $\xiov\sim |\eps|^{-1/\alpha}$ will be observed only if $\xiov$ is sufficiently large (i.e., $|\eps|$ sufficiently small). Thus, to determine the chaos exponent from the numerical solution of the flow equations, one would need both a very small value of $\eps$ and an extremely large $\nmax$, which cannot be realized in practice.

\subsubsection{Comparison with Duemmer and Le Doussal's work}
\label{subsubsec_DLD} 

Equations~(\ref{rgeq1},\ref{rgeq2}) and (\ref{linear4}) are identical to those obtained by DLD in their study of periodic elastic manifolds pinned by disorder (with the temperature playing the role of the Luttinger parameter).\cite{Duemmer07} But our analysis of the linearized equation~(\ref{linear4}) differs in a crucial way: DLD use $\delta^*{}''(u)=1/\pi\bar l_2$ which contradicts~(\ref{deltafppp}) and is not correct since it violates the condition $\int_0^\pi du\, \delta^*{}''(u)=0$. As a result, they obtain equation~(\ref{linear6}) for the function $g$ instead of $f=g-\pi\sh$.\cite{not7} Since none of the solutions of this equation have a vanishing integral over the interval $[0,1]$ they conclude that the linear analysis of perturbations about the fixed point fails. 

To circumvent this difficulty DLD consider a two-dimensional system where the temperature is marginal, $\theta=0$, and therefore does not flow under RG. When $\eps=0$ one obtains a line of fixed points indexed by $T>0$. The function $\delta^*(u)$ is  analytic and exhibit a thermal boundary layer (TBL) of width $T$ instead of the cusp, which makes the linear analysis about the fixed point free of the difficulties that arise when the temperature flows toward zero. DLD find that outside the TBL (i.e. for $|u|\gtrsim T$) the eigenfunction $g(u)$ corresponding to the largest eigenvalue $\lamb$ must be chosen among the solutions $f(u)$ of~(\ref{linear6}), whereas\cite{not2}
\beq 
g(u) = \frac{1}{6T} \frac{\lambda-2}{\left[1+\left(\frac{u}{2T}\right)^2\right]} 
\label{DLD1}
\eeq 
for $|u|\lesssim T$. The eigenvalue 
\beq 
\lamb \simeq 2 - \frac{3}{\ln(1/T)}
\label{DLD2}
\eeq
converges logarithmically toward 2 when $T\to 0$. We conclude that the $T\to 0$ limit of DLD's results in the marginal case $\theta=0$ agree with the conclusions of Sec.~\ref{subsubsec_CBL_linear} obtained in the case where the temperature $T$ (i.e., the Luttinger parameter in our notations) flows to zero ($\theta>0$). Our results show that the $T\to 0$ limit of the function $g(u)$ inside the TBL [Eq.~(\ref{DLD1})] is given by the singular function $\sh(u)$. The temperature dependence of the eigenvalue in~(\ref{DLD2}) is similar to the dependence of $\lambda$ with respect to $\nmax$ in~(\ref{lambvsnmax}). These logarithmic corrections are due to the finite length scale introduced by the finite temperature in the marginal case ($\theta=0$) or the finite number of circular harmonics in our study ($\theta>0$). 

DLD also consider a system with dimensionality larger than two where the temperature is irrelevant ($\theta>0$). In that case they find that the escape from the fixed point, i.e., the growth of $\delta_{12,k}(u)-\delta_{ii,k}(u)$, occurs with eigenvalue $\lambda=2$ if $u$ is outside the CBL and $2+\theta$ if $u$ is inside the CBL, which gives the chaos exponent $\alpha=2/(2+\theta)$. This latter result disagrees with our conclusions.

\section{Quantum chaos} 
\label{sec_qchaos} 

In this section we consider two copies of the system subjected to the same disorder potential but with different Luttinger parameters, 
\beq
\begin{split}
	K_1 &= K + \eps, \\
	K_2 &= K - \eps .
\end{split}
\eeq 
The replicated action is then given by  
\begin{multline}
S[\{\varphi_{ia}\}] = \sum_{i,a} \int_{x,\tau} \frac{v}{2\pi K_i} \left\{ (\dx\varphi_{ia})^2 + \frac{(\dtau\varphi_{ia})^2}{v^2} \right\} \\
- \sum_{a,b,i,j} \calD \int_{x,\tau,\tau'} \cos[2\varphi_{ia}(x,\tau) - 2\varphi_{jb}(x,\tau')] .
\label{action1} 
\end{multline} 
In order to implement the FRG, we choose a cutoff function which depends on the copy index, 
\beq 
R_{i,k}(q,i\w) = Z_{ix}  \left(q^2 + \frac{\w^2}{v_{i,k}^2} \right) r\left( \frac{q^2+\w^2/v_{i,k}^2}{k^2} \right) , 
\eeq  
where $Z_{ix}=v/\pi K_i$ and $v_{i,k}$ is the renormalized velocity of the $i$th copy. 

The ansatz for the effective action $\Gamma_k[\phiset]$ is given by~(\ref{ansatz2}) with $Z_x$ and $v_k$ replaced by $Z_{ix}$ and $v_{i,k}$. The flow equations for $K_{i,k}$, $v_{i,k}$ and 
\beq 
\delta_{ij,k}(u) = - \frac{K_i K_j}{v^2 k^3} V''_{ij,k}(u)  
\eeq
are given by 
\begin{equation} 
\begin{split}
\dt K_{i,k} &= \theta_{i,k} K_{i,k} , \qquad
\dt (K_{i,k}/v_{i,k}) = 0 , \\
\theta_{i,k} &= \frac{\pi}{2} \delta''_{ii,k}(0) \bar m_\tau . 
\end{split}
\end{equation}
and 
\begin{align} 
\dt\delta_{ii,k}(u) ={}& -3 \delta_{ii,k}(u) - l_1 K_{i,k} \delta''_{ii,k}(u) \nonumber \\ & \hspace{-1.5cm}
+ \pi \bar l_2 \{ \delta_{ii,k}''(u) [\delta_{ii,k}(u)-\delta_{ii,k}(0)] + \delta'_{ii,k}(u)^2 \} ,  \label{rgeq11} \\ 
\dt\delta_{12,k}(u) ={}& -3 \delta_{12,k}(u)  - l_1 \frac{K_{1,k}+K_{2,k}}{2}  \delta''_{12,k}(u)  \nonumber \\ &  \hspace{-1.5cm}
+ \pi \bar l_2 \{ \delta_{12,k}''(u) [\delta_{12,k}(u)-\delta_{12,k}(0)] + \delta'_{12,k}(u)^2 \} \nonumber \\ &  - \pi \bar l_2 \hat K_k \delta''_{12}(u), \label{rgeq22} 
\end{align}
where 
\beq 
\hat K_k = \frac{\delta_{11,k}(0)+\delta_{22,k}(0)}{2} - \delta_{12,k}(0) .
\eeq 
These equations are similar to those discussed in Secs.~\ref{sec_model} and \ref{sec_dchaos}. Since $\hat K_\Lambda=\calO(\epsilon^2)$, all conclusions reached in Sec.~\ref{sec_dchaos} remain valid as can be explicitly verified by solving numerically the flow equations. The chaotic behavior now originates in the difference in the quantum fluctuations of the two copies of the system. Although they are suppressed in the long-distance limit, $K_{i,k}\to 0$ for $k\to 0$, they select different ground states in the two copies. This ``quantum'' chaos is analog to the ``temperature'' chaos in classical disordered system, as shown by the analogy between Eqs.~(\ref{rgeq11}) and (\ref{rgeq22}) and the equations derived in Ref.~\onlinecite{Ledoussal06}.

\section{Conclusion} 

We have investigated the chaotic behavior of the BG phase of a one-dimensional disordered Bose fluid. By solving numerically the nonperturbative FRG equations, we find that two copies of the system with slightly different disorder configurations become statistically uncorrelated at large distances. The chaos exponent $\alpha$ can be obtained from the overlap length $\xiov\sim |\eps|^{-1/\alpha}$ or the growth of $\delta_{12,k}(0)-\delta^*(0)\sim e^{-2\alpha t}$ at long RG time $|t|$, but the convergence with the number $\nmax$ of circular harmonics used for the disorder correlators $\delta_{ij,k}(u)$ turns out to be logarithmic and therefore extremely slow. From the linear analysis of perturbations about the BG fixed point, we are however able to show analytically that $\alpha=1$. 

Although the chaos exponent is related to the relevant RG eigenvalue $\lambda_{\rm max}=2\alpha$ of the linearized flow near the BG fixed point, as for a standard critical point, the peculiar nature of the fixed point makes the situation somewhat unusual. The fixed-point disorder correlator $\delta^*(u)$ exhibits cusps at $u=p\pi$ and $\delta_{12,k}(u)$ approaches and departs from its nonanalytic fixed-point form {\it via} a QBL and a CBL, respectively. This has strong consequences for the linear analysis of the perturbations about the fixed point. The eigenfunctions $g_{12,k}(u)\equiv \delta_{12,k}(u)-\delta^*(u)=\pi\sh(u)+f(u)$, solutions of the linearized flow equations, are singular at $u=p\pi$ ($f(u)$ is a regular function). Although this could call into question the linear analysis, the agreement with the results obtained from the numerical analysis of the flow equations, where the function $\delta_{12,k}(u)$ remains analytic at all scales $k\geq 0$, strongly supports its validity, even if this agreement is obtained for a finite number $\nmax$ of circular harmonics for which the chaos exponent significantly differs from its converged value.

The chaotic behavior of the BG phase can also be induced by a modification of quantum fluctuations due to a slight variation of the Luttinger parameter. 

Finally we note that all these conclusions also apply to the Mott-glass phase of a disordered Bose fluid induced by long-range interactions.\cite{Daviet20}

\begin{acknowledgments} 
	We thank Gilles Tarjus for a critical reading of the manuscript. 
\end{acknowledgments} 
 
%


\begin{thebibliography}{41}
\bibitem [{not({\natexlab{a}})}]{not6}%
	\BibitemOpen
	\href@noop {} {}  \bibinfo {note} {In the single-copy case,
		the nonperturbative FRG approach has also been used to study the random-field
		Ising model, see
		Refs.~\onlinecite{Tarjus08,Tissier08,Tissier12,Tissier12a,Tarjus19}.}\BibitemShut
	{Stop}%
\bibitem [{not({\natexlab{b}})}]{not3}%
	\BibitemOpen
	\href@noop {} {}  \bibinfo {note} {The boundary layer, of
		width $\sim K_k$, implies that $\delta_k(0)-\delta^*(0)\sim K_k$ to leading
		order in the limit $K_k\to 0$.}\BibitemShut {Stop}%
\bibitem [{not({\natexlab{c}})}]{not8}%
	\BibitemOpen
	\href@noop {} {} \bibinfo {note} {The $\pi$-periodic
		Kronecker and Dirac combs are defined by
		$\sh^{(K)}_{u,0}=\sum_{n=-\infty}^\infty \delta^{(K)}_{u,n\pi}$ and
		$\sh(u)=\sum_{n=-\infty}^\infty \delta(u-n\pi)$, respectively
		($\delta^{(K)}_{u,0}$ denotes the Kronecker delta).}\BibitemShut {Stop}%
\bibitem [{not({\natexlab{e}})}]{not1}%
	\BibitemOpen
	\href@noop {} {} \bibinfo {note} {Note that even if
		$h''(u)$ contains Dirac peaks, $h''(u) [\delta^*(u)-\delta^*(0)]$ is a
		regular function since $\delta^*(u)-\delta^*(0)$ vanishes for
		$u=p\pi$.}\BibitemShut {Stop}%
\bibitem [{not({\natexlab{g}})}]{not2}%
	\BibitemOpen
	\href@noop {} {} \bibinfo {note} {When comparing with
		DLD's work, we have set the dimensionality to one (i.e., $\eps=3$ in DLD's
		notations) in the results of Ref.~\onlinecite{Duemmer07}.}\BibitemShut
	{Stop}%
\end{thebibliography}

\end{document}